# Relaxation Penalties and Priors for Plausible Modeling of Nonidentified Bias Sources

**Sander Greenland**

*Abstract.* In designed experiments and surveys, known laws or design feat ures provide checks on the most relevant aspects of a model and identify the target parameters. In contrast, in most observational studies in the health and social sciences, the primary study data do not identify and may not even bound target parameters. Discrepancies between target and analogous identified parameters (biases) are then of paramount concern, which forces a major shift in modeling strategies. Conventional approaches are based on conditional testing of equality constraints, which correspond to implausible point-mass priors. When these constraints are not identified by available data, however, no such testing is possible. In response, implausible constraints can be relaxed into penalty functions derived from plausible prior distributions. The resulting models can be fit within familiar full or partial likelihood frameworks.

The absence of identification renders all analyses part of a sensitivity analysis. In this view, results from single models are merely examples of what might be plausibly inferred. Nonetheless, just one plausible inference may suffice to demonstrate inherent limitations of the data. Points are illustrated with misclassified data from a study of sudden infant death syndrome. Extensions to confounding, selection bias and more complex data structures are outlined.

*Sander Greenland is Professor, Departments of Epidemiology and Statistics, University of California, Los Angeles, California 90095-1772, USA e-mail: lesdomes@ucla.edu.*

## 1. BACKGROUND

### 1.1 Observational Epidemiologic Data Identify Nothing

With few exceptions, observational data in the health and social sciences identify no parameter whatsoever unless assumptions of uncertain status are made (Greenland, 2005a). Even so-called "nonparametric identification" depends on assumptions that are only working hypotheses, such as absence of uncontrolled bias. Worse, close inspection of the actual processes producing the data usually reveals far more complexity than can be fully modeled in a reasonable length of time. Formal analyses, no matter how mathematically sound and elegant, never fully capture the uncertainty warranted for inferences from such data, and in epidemiology and medicine often lead to inferences that are later judged much farther off target than random error alone could explain (Lawlor et al., 2004).

Many of the issues can be seen in simple cases. Suppose our target parameter is the prevalence $\Pr(T = 1)$ in a target population of a health-related exposure indicator $T$, to be estimated from a sample of $N$ persons for whom $T$ is measured. If $A$ is the number of sampled persons with $T = 1$, the conventional binomial model leads to an unbiased estimator $A/N$ of $\Pr(T = 1)$ and many procedures for constructing interval estimates. But $N$ is often considerably less than the number of eligible persons for whom contact attempts were made, leaving uncertainty about what subset of the target was actually sampled. One may not even be able to assume independent responses: Physicians may refuse to provide their entire block of patients, or patients may encourage friends to participate, and these actions may be related to $T$. Thus, the binomial model for $A$ is a convention adopted uncritically from simple and complete random surveys; there are many ways this model may fail, none testable from the data $(A, N)$ alone.

Next, suppose we have only an imperfect measure $X$ of $T$ (e.g., a self-report of $T$) in the sample. The observable variable is now the count $A$ of $X = 1$, and binomial inference on $\Pr(T = 1)$ from $A/N$ alone is unsupported even with random sampling. Yet the usual if implicit convention is to pretend that $\Pr(X = 1) = \Pr(T = 1)$ and discuss the impact of violations intuitively (e.g., Fortes et al., 2008). This is a poor strategy because the convention derives from the assumption $X = T$, which no informed observer holds, and anchors subsequent intuitions to this extreme case. The problem can be addressed by obtaining error-free measurements of $T$, but that is often impossible, for example, when $T$ represents a lifetime chemical exposure or nutrient intake. At best we might obtain alternate measurements of $T$ and incorporate them into a predictive model for $T$, which would have its own nonidentified features.

### 1.2 Identification versus Plausibility

To summarize the problem, conventional statistics are derived from design mechanisms (such as random sampling) or known physical laws that enforce the assumed data model; but studies based on passive observation of health and social phenomena (such as studies of health-care data bases) have little or nothing in the way of such enforcement, leaving us no assurance that conventional statistics (even when nonparametric) are estimating the parameter of interest. Furthermore, because the actual data-generating process depends on latent variables related to the target parameter, that parameter is not identified from the observed data. These studies thus suffer from a curse of nonidentification, in that identification can be achieved only by adding constraints that are neither enforced by known mechanisms nor testable with the observed data.

In light of this problem, many epidemiologic authors have emphasized the need to unmoor observational data analysis from conventional anchors (Phillips, 2003; Lash and Fink, 2003; Maldonado, 2008). There is now a vast literature on models to fulfill this need, sometimes described under the general heading of bias analysis (Greenland, 2003a, 2005a, 2009; Greenland and Lash, 2008). Examples include models for selection biases (Copas, 1999; Geneletti et al., 2009; Scharfstein, Rotnitsky and Robins, 1999, 2003), nonignorable missingness and treatment assignment (Kadane, 1993; Baker, 1996; Molenberghs, Kenward and Goetghebeur, 2001; Little and Rubin, 2002; Robins, Rotnitzky and Scharfstein, 2000; Rosenbaum, 2002; Vansteelandt et al., 2006), uncontrolled or collinear confounders (Bross, 1967; Leamer, 1974; Greenland, 2003a; Gustafson and Greenland, 2006; McCandless et al., 2007; Yanagawa, 1984), measurement error (Gustafson, 2003, 2005; Greenland, 2009) and multiple biases (Eddy, Hasselblad and Shachter, 1992; Greenland, 2003a, 2005a; Molitor et al., 2008; Goubar et al., 2008; Turner et al., 2009; Welton et al., 2009).

Despite the profusion of literature on the topic, integration of core ideas of bias analysis into basic education, software and practice has been slow. One obstacle may be the diversity of approaches proposed. Another may be the failure to connect them to familiar, established methods. Yet another obstacle may be the greater demand for contextual input that most require. Central to that input is the informal but crucial concept of a *plausible* model. I will call a model plausible if it appears to neither conflict with accepted facts nor assume far more facts than are in evidence. Implausible models are then models rejectable a priori as either conflicting with or going too far beyond existing background information.

The distinction between plausible and implausible models is fuzzy, shifting and disputable, but many models will clearly be implausible within a given context. For example, models that assume zero exposure measurement error ($X = T$ above) are very implausible in environmental, occupational and nutritional epidemiology, because no one can plausibly argue such errors are absent. In fact, most conventional data-probability models appear implausible in epidemiologic contexts. Such models are often rationalized as providing data summaries about identified parameters such as $\Pr(X = 1)$ above, but their outputs are invariably interpreted as inferences about targets such as $\Pr(T = 1)$. Avoiding such misinterpretations requires model expansion into the nonidentified dimensions that connect observables to targets.

Plausibility concepts apply to models for prior probabilities as well as to models for data-generating processes. For example, consider a prior for a disease prevalence $\pi$ that assigned $\Pr(\pi = 0.5) = 0.5$ and was uniform over the rest of the unit interval. This prior would be implausible as an informed opinion because no genuine epidemiologic evidence could provide such profound support for $\pi = 0.5$ and yet fail to distinguish among every other possibility. Analogous criticisms apply to most applications of "spike-and-slab" priors in the health and social sciences.

### 1.3 Outline of Paper

The present article reviews the above points, focusing on plausible extensions of conventional models in order to simplify bias analysis for teaching and facilitate its conduct with ordinary software. It begins by outlining a likelihood-based framework for observational data analysis that mixes frequentist and Bayesian ideas (as has long been recommended, e.g., Box, 1980; Good, 1983). It stands in explicit opposition to the notion that use of priors demands a fully Bayesian framework or exact posterior computation, even though partial priors in some form are essential for inferences on nonidentified parameters. Instead, it encourages use of partial priors as identifying penalty functions. These functions may be translated into augmenting data, which aids plausibility evaluation and facilitates computation with familiar likelihood and estimating-equation software.

Section 3 illustrates points with data from a large collaborative case-control study of sudden infant death syndrome (SIDS). It starts with conventional analyses of the data, describes a misclassification problem, then provides analyses using priors only for nonidentified parameters. Section 4 then outlines extensions to "validation" data, and describes how the misclassification model can be re-interpreted to handle uncontrolled confounders and selection bias. Throughout, the settings of concern are those in which the data have been collected but a "correct" model for their generation can never be known or even approximated. In these settings, we cannot even guarantee that inferences from the posterior will be superior to inferences from the prior (Neath and Samaniego, 1997). Thus, the importance of specific models and priors is de-emphasized in favor of providing a framework for sensitivity analysis across plausible models and priors. This framework need not be all-encompassing, because often just a few plausible specifications can usefully illustrate the illusory nature of an apparently conclusive conventional analysis.

## 2. PRIORS AND PENALTIES AS TOOLS FOR ENHANCING MODEL PLAUSIBILITY

### 2.1 Models and Constraints

The formalism used here is similar to that in Greenland (2005a) and Vansteelandt et al. (2006), tailored to a profile penalized-likelihood approach. Consider a family of models $\mathbf{G} = \{G(\mathbf{a}; \boldsymbol{\gamma}, \boldsymbol{\theta}) : (\boldsymbol{\gamma}, \boldsymbol{\theta}) \in \boldsymbol{\Gamma} \times \boldsymbol{\Theta}\}$ for the distribution of an observable-data structure $\mathbf{A}$ taking values $\mathbf{a}$ in a sample space $\mathcal{A}$ with $\mathbf{G}$ satisfying any necessary regularity conditions. The inferential target parameter will be a function $\boldsymbol{\tau} = \boldsymbol{\tau}(\boldsymbol{\gamma}, \boldsymbol{\theta})$ of the model parameters.

$\mathbf{G}$ represents a set of constraints on an unknown objective frequency distribution or "law" for $\mathbf{A}$. In classical applied statistics these constraints are induced by study design or physical laws. In contrast, in observational health and social sciences these constraints are largely or entirely hypothetical, which motivates the present treatment. The separation of the total parameter $(\boldsymbol{\gamma}, \boldsymbol{\theta})$ into components $\boldsymbol{\gamma}$ and $\boldsymbol{\theta}$ is intended to reflect some conceptual distinction that drives subsequent analyses and will be clarified below. The assumption $(\boldsymbol{\gamma}, \boldsymbol{\theta}) \in \boldsymbol{\Gamma} \times \boldsymbol{\Theta}$ (variation independence of $\boldsymbol{\gamma}$ and $\boldsymbol{\theta}$) provides technical simplifications when using partial priors (Gelfand and Smith, 1999) and will be discussed in Section 3.7.

When $(\boldsymbol{\gamma}, \boldsymbol{\theta})$ is identified, $\boldsymbol{\gamma}$ could contain the parameters considered essential to retain in the model, whereas $\boldsymbol{\theta}$ could contain parameters considered optional, as when $\boldsymbol{\theta}$ contains regression coefficients of candidate variables for deletion. Conventional modeling then considers only the following:

(1) Equality constraints (also known as "hard," "sharp" or "point" constraints) of the form $r(\boldsymbol{\theta}) = \mathbf{c}$, where $\mathbf{c}$ is a known constant (usually $\mathbf{0}$). This reduces the model family to $\{G(\mathbf{a}; \boldsymbol{\gamma}, \boldsymbol{\theta}) : (\boldsymbol{\gamma}, \boldsymbol{\theta}) \in \boldsymbol{\Gamma} \times r^{-1}(\mathbf{c})\}$, where $r^{-1}(\mathbf{c})$ is the preimage of $\mathbf{c}$ in $\boldsymbol{\Theta}$, and constrains $\boldsymbol{\tau}$ to $\{\boldsymbol{\tau}(\boldsymbol{\gamma}, \boldsymbol{\theta}) : (\boldsymbol{\gamma}, \boldsymbol{\theta}) \in \boldsymbol{\Gamma} \times r^{-1}(\mathbf{c})\}$.

(2) No constraint apart from logical bounds (e.g., 0 and 1 for a probability): both $\boldsymbol{\gamma}$ and $\boldsymbol{\theta}$ are treated as "unknown constants," which corresponds to no constraint on the target parameter $\boldsymbol{\tau} = \boldsymbol{\tau}(\boldsymbol{\gamma}, \boldsymbol{\theta})$.

The choice between these extremes is usually based on a test of the constraint $r(\boldsymbol{\theta}) = \mathbf{c}$, often derived from the likelihood function $L(\boldsymbol{\gamma}, \boldsymbol{\theta}; \mathbf{a}) = G(\mathbf{a}; \boldsymbol{\gamma}, \boldsymbol{\theta})$ when $\mathbf{G}$ is an exponential family, for example, by contrasting the maximum of the deviance $-2L(\boldsymbol{\gamma}, \boldsymbol{\theta}; \mathbf{a})$ with and without the constraint.

In the problems considered in the present paper, options (1) and (2) are not available because $\boldsymbol{\theta}$ is not identified, in the sense that, for each $\mathbf{a} \in \mathcal{A}$, the profile likelihood $L(\boldsymbol{\theta}; \mathbf{a}) \equiv \max_{\boldsymbol{\gamma} \in \boldsymbol{\Gamma}} L(\boldsymbol{\gamma}, \boldsymbol{\theta}; \mathbf{a})$ is constant. Thus, no test of $r(\boldsymbol{\theta}) = \mathbf{c}$ is available without introducing other nonidentified constraints. Consider again the misclassification example observing $A = a$ for the $X = 1$ count, with $\tau = \Pr(T = 1)$ and $\theta = \Pr(X = 1) - \Pr(T = 1)$. Then $L(\tau, \theta; a) \propto (\tau + \theta)^a (1 - \tau - \theta)^{N-a}$ with $L(\theta; a) = (a/N)^a (1 - a/N)^{N-a}$, a constant; thus, we cannot test $\theta = 0$ to evaluate use of $X$ for $T$ in inference about $\tau$. In fact, we may reparameterize to remove $\theta$ from the likelihood: Defining $\gamma = \Pr(X = 1)$, we obtain $\Pr(T = 1) = \tau = \gamma - \theta$ and $L(\gamma, \theta; a) \propto \gamma^a (1 - \gamma)^{N-a}$, a *transparent* parameterization (Gustafson, 2005). This parameterization shows that observation a places no constraint on $\theta$ and hence no constraint on the target parameter $\tau = \gamma - \theta$. Thus, $\tau$ is not even partially identified, despite having an identified component $\gamma$.

## 2.2 Sensitivity to Bias Parameters

Because $\boldsymbol{\theta}$ determines the discrepancy between the target $\boldsymbol{\tau}$ and the identified parameter $\boldsymbol{\gamma}$ often estimated as if it were the target, $\boldsymbol{\theta}$ may be called a *bias parameter* (Greenland, 2005a). Because inferences that are sensitive to nonidentified parameters will remain asymptotically sensitive to constraints on those parameters, $\boldsymbol{\theta}$ has also been called a *sensitivity* parameter (Moleberghs, Kenward and Goetghebeur, 2001). Conventional sensitivity analysis shows how inferences change as equality constraints are varied, for example, as $\mathbf{c}$ in $\boldsymbol{\theta} = \mathbf{c}$ is varied (Rosenbaum, 2002; Greenland and Lash, 2008).

Vansteelandt et al. (2006) allow relaxation of such point constraints into a constraint of the form $\boldsymbol{\theta} \in \mathbf{R}$, where $\mathbf{R}$ represents a plausible range for $\boldsymbol{\theta}$. This constraint may be written as $r(\boldsymbol{\theta}) = 1$, where $r(\boldsymbol{\theta})$ is the membership indicator for $\mathbf{R}$. Let $\boldsymbol{\gamma}_0$ be the true value of $\boldsymbol{\gamma}$, which is unknown but identified. Assuming $\boldsymbol{\theta} \in \mathbf{R}$ then constrains $\boldsymbol{\tau}$ to $\{\boldsymbol{\tau}(\boldsymbol{\gamma}, \boldsymbol{\theta}) : (\boldsymbol{\gamma}, \boldsymbol{\theta}) \in \boldsymbol{\Gamma} \times \mathbf{R}\}$ and identifies the set $\{\boldsymbol{\tau}(\boldsymbol{\gamma}_0, \boldsymbol{\theta}) : \boldsymbol{\theta} \in \mathbf{R}\} = \boldsymbol{\tau}(\boldsymbol{\gamma}_0, \mathbf{R})$. Vansteelandt et al. call $\boldsymbol{\tau}(\boldsymbol{\gamma}_0, \mathbf{R})$ an *ignorance region* for $\boldsymbol{\tau}$ and propose frequentist estimators of this region as a means of summarizing a sensitivity analysis in which $\boldsymbol{\theta}$ is varied over $\mathbf{R}$. To illustrate with the misclassification example, the constraint $|\theta| < 0.2$ corresponds to $\theta \in \mathbf{R} = [0, 0.2)$ and identifies the ignorance region $\boldsymbol{\tau}(\gamma_0, \mathbf{R}) = \{|\gamma_0 - \theta| : |\theta| < 0.2\}$.

From a Bayesian perspective, sensitivity analysis by varying $\mathbf{c}$ in $r(\boldsymbol{\theta}) = \mathbf{c}$ can be viewed as analyses of sensitivity to priors in which the priors are limited to point masses $\Pr(r(\boldsymbol{\theta}) = \mathbf{c}) = 1$. Similarly, analyses based on $\boldsymbol{\theta} \in \mathbf{R}$ can be viewed as using a prior restricted to $\mathbf{R}$. Why limit analyses to such sharply bounded constraints or priors? A practical argument might be that there are too many possible constraints or priors and, thus, some limit on their form is needed. But typical equality constraints (point priors) are very implausible, insofar as they make assertions far beyond that warranted by available evidence; that is, they are much too informative. Similarly, restricting $\boldsymbol{\theta}$ to a sharply bounded region $\mathbf{R}$ risks completely excluding values of $\boldsymbol{\theta}$ that are plausible (and perhaps correct); expansion of $\mathbf{R}$ to avoid this risk may result in a practically uninformative region $\boldsymbol{\tau}(\boldsymbol{\gamma}_0, \mathbf{R})$ for $\boldsymbol{\tau}$. A broad region $\mathbf{R}$ also ignores what may be substantial differences in plausibility among its members.

## 2.3 Relaxation Penalties and Priors

To address the deficiencies of point constraints for $\boldsymbol{\theta}$, we may instead relax (expand) the constraints into a family $\mathbf{D} = \{D(\boldsymbol{\theta}; \boldsymbol{\lambda}) : \boldsymbol{\lambda} \in \boldsymbol{\Lambda}\}$ of penalty functions indexed by $\boldsymbol{\lambda}$ which subsumes the point constraints as special or limiting cases. This situation arises in scatterplot smoothing, where $\boldsymbol{\gamma}$ may contain an intercept, linear and quadratic term and $\boldsymbol{\theta}$ may contain distinct cubic terms for each design point, thus leaving $(\boldsymbol{\gamma}, \boldsymbol{\theta})$ nonidentified. Point constraints (such as setting all nonlinear terms to zero) exclude entire dimensions of the regression space, and hence risk oversmoothing. In contrast, having no constraint gives back the raw data points as the fitted curve, resulting in no smoothing. Penalties provide "soft" or "fuzzy" constraints that relax the sharp constraints of conventional models to produce smooth curves between these extremes (Hastie and Tibshirani, 1990).

Penalization is a form of shrinkage estimation, wherein asymptotic unbiasedness may be sacrificed in exchange for reduced expected loss. Nonetheless, in identified models mild penalties can also reduce asymptotic bias relative to ordinary maximum likelihood (Bull, Lewinger and Lee, 2007). In observational studies the potential gain from penalization is far greater because unbiasedness can be derived only by assuming greatly oversimplified models that are likely biased (Greenland, 2005a). In particular, an estimator unbiased under a model $G(\mathbf{a}; \boldsymbol{\gamma}, \mathbf{c})$ may suffer enormous bias if the point constraint $\boldsymbol{\theta} = \mathbf{c}$ is incorrect. Penalties that relax $\boldsymbol{\theta} = \mathbf{c}$ to a weaker form can reduce this source of bias (Greenland, 2000; Gustafson and Greenland, 2006, 2010), although unbiasedness is arguably an unrealistic goal in these settings.

Given $\mathbf{D}$, popular strategies for choosing $\boldsymbol{\lambda}$ include empirical-Bayes and cross-validation (Hastie, Tibshirani and Friedman, 2001). In the present applications, however, $\boldsymbol{\lambda}$ is not identified and so external grounds for choosing $\boldsymbol{\lambda}$ are needed. The interpretation of a penalty function $D(\boldsymbol{\theta}; \boldsymbol{\lambda})$ as the transform $-2\ln\{H(\boldsymbol{\theta}; \boldsymbol{\lambda})\}$ of a prior density $H(\boldsymbol{\theta}; \boldsymbol{\lambda})$ can provide contextual guidance for making good choices. To illustrate, let $\boldsymbol{\lambda} = (\boldsymbol{\mu}, \mathbf{W})$ with $\mathbf{W}$ a positive-definite information matrix. Then the quadratic (generalized ridge-regression) penalty $(\boldsymbol{\theta} - \boldsymbol{\mu})'\mathbf{W}(\boldsymbol{\theta} - \boldsymbol{\mu})$ corresponds to a normal$(\boldsymbol{\mu}, \mathbf{W}^{-1})$ prior on $\boldsymbol{\theta}$ (Titterington, 1985). For diagonal $\mathbf{W}$ the absolute (Lasso) penalty $|\boldsymbol{\theta} - \boldsymbol{\mu}|'\mathbf{w}^{1/2}$ corresponds to independent double-exponential (Laplacian) priors with mean and scale vectors $\boldsymbol{\mu}$ and $\mathbf{w}^{-1/2}$ where $\mathbf{w}$ is the diagonal of $\mathbf{W}$ (Tibshirani, 1996).

Taking $\boldsymbol{\mu} = \mathbf{c}$, in either case the point constraint $\boldsymbol{\theta} = \mathbf{c}$ is now the limiting penalty as $\mathbf{W}^{-1}$ goes to $\mathbf{0}$, and thus corresponds to infinite information. We should instead want to choose $\mathbf{W}$ such that the resulting $H(\boldsymbol{\theta}; \boldsymbol{\lambda})$ is no more informative than we find plausible and assigns more than negligible odds (relative to the maximum) to all plausible possibilities.

The form of the resulting penalty will allow varying degrees of plausibility over $\boldsymbol{\Theta}$, as it should; $\mathbf{c}$ becomes the most plausible value. This use of priors to relax sharp constraints on nonidentified parameters does not entail commitment to Bayesian philosophy, since the resulting penalized estimators can still be evaluated based in part on their frequency properties (Gustafson, 2005; Gustafson and Greenland, 2006, 2010).

In general, interpretation of a given penalty $D(\boldsymbol{\theta}; \boldsymbol{\lambda})$ involves transformation to see which if any $\boldsymbol{\lambda}$ yield contextually reasonable prior densities $H(\boldsymbol{\theta}; \boldsymbol{\lambda}) \propto \exp(-D(\boldsymbol{\theta}; \boldsymbol{\lambda})/2)$. If $\exp(-D(\boldsymbol{\theta}; \boldsymbol{\lambda})/2)$ is not integrable, $D(\boldsymbol{\theta}; \boldsymbol{\lambda})$ will not completely identify $\boldsymbol{\theta}$ (the implied prior is improper), although $\boldsymbol{\tau}$ or some lower-dimensional function of it may still have a proper prior (Gelfand and Sahu, 1999). One may also vary $\boldsymbol{\lambda}$ to assess sensitivity to its choice, or give $\boldsymbol{\lambda}$ a prior as in Bayes empirical-Bayes estimation (Deely and Lindley, 1981; Samaniego and Neath, 1996).

## 2.4 Partial Priors

For a pragmatic frequentist who uses priors to impose penalties or soft constraints, use of a prior for $\boldsymbol{\theta}$ alone is natural. For a pragmatic Bayesian, placing a prior on $\boldsymbol{\theta}$ alone is an effort-conserving strategy, recognizing that thorough exploration of all priors is infeasible and that the dimensions demanding greatest care in prior specification are the nonidentified ones (Neath and Samaniego, 1997). In contrast, some or all identified dimensions may be judged not worth the effort of formalizing, especially if the data have enough information in those dimensions to overwhelm any cautious or vague prior.

With prior specification limited to $\boldsymbol{\theta}$, the implicit prior $p(\boldsymbol{\gamma}, \boldsymbol{\theta}) = p(\boldsymbol{\theta})$ is improper on $\boldsymbol{\Gamma}$, with posterior $p(\boldsymbol{\gamma}, \boldsymbol{\theta}|\mathbf{a}) \propto L(\boldsymbol{\gamma}, \boldsymbol{\theta}; \mathbf{a})H(\boldsymbol{\theta}; \boldsymbol{\lambda})$. The log posterior is then a loglikelihood for $(\boldsymbol{\gamma}, \boldsymbol{\theta})$ with penalty $-2\ln\{H(\boldsymbol{\theta}; \boldsymbol{\lambda})\}$. The resulting penalized-likelihood analyses have been called partial-Bayes or semi-Bayes (Cox, 1975; Greenland, 1992; Bedrick, Christensen and Johnson, 1996); they can also be applied if $L(\boldsymbol{\gamma}, \boldsymbol{\theta}; \mathbf{a})$ is a partial, pseudo or weighted likelihood. Conventional analyses are extreme cases in which $\boldsymbol{\theta}$ is given a point prior. Identification of $(\boldsymbol{\gamma}, \boldsymbol{\theta})$ by $L(\boldsymbol{\gamma}, \boldsymbol{\theta}; \mathbf{a})H(\boldsymbol{\theta}; \boldsymbol{\lambda})$ corresponds to a unique maximum penalized-likelihood estimate (MPLE) and a proper posterior distribution for $(\boldsymbol{\gamma}, \boldsymbol{\theta})$.

## 2.5 Plausible Penalties and Data Priors

Not all penalties or priors will appear plausible. One way to evaluate plausibility is to construct a thought experiment with sample space $\mathcal{B}$ such that $H(\theta; \lambda)$ is the profile likelihood $L(\theta; \mathbf{b}_\lambda, \lambda)$ for $\theta$ derived from an outcome $\mathbf{b}_\lambda \in \mathcal{B}$. Specifically, we examine a family $\mathbf{F} = \{F(\mathbf{b}; \theta, \lambda, \delta), (\theta, \delta) \in \Theta \times \Delta\}$ of distributions conjugate to the prior-distribution family $\mathbf{H}$ in that the "data prior" $\mathbf{b}_\lambda$ yields $L(\theta; \mathbf{b}_\lambda, \lambda) = \max_{\delta \in \Delta} F(\mathbf{b}_\lambda; \theta, \lambda, \delta) = H(\theta; \lambda)$ (Higgins and Spiegelhalter, 2002; Greenland, 2003b, 2007a, 2007b, 2009); $\delta$ contains any nuisance parameters in the chosen experiment. The experiment and its parameterization is chosen to make $\delta$ variation independent of $\gamma$; there may, however, be no need for $\delta$ and so it will henceforth be dropped.

To illustrate, consider again the binomial-survey example reparameterized to $\gamma = \text{logit}\{\Pr(X = 1)\}$ and $\text{logit}\{\Pr(T = 1)\} = \gamma - \theta$, so that $\theta$ now represents the asymptotic bias in $\text{logit}(A/N)$ as an estimator of $\text{logit}\{\Pr(T = 1)\}$. A convenient prior family for $\theta$ is the generalized-conjugate or log-$F$ distribution (Greenland, 2003a, 2003b, 2003c; Jones, 2004), which has density $H(\theta; \lambda) \propto e^{znr}/(1 + e^z)^n$ where $z = (\theta + \text{logit}(r) - m)/s$ and $\lambda = [m, s, r, n]'$; $m$ and $s$ are the desired mode and scale for the prior, $0 < r < 1$ controls skewness, and $n > 0$ controls tail weight (thinner tails as $n$ increases). When $r = 0.5$ this $H(\theta; \lambda)$ is symmetric; it then equals the logistic density when $n = 2$, and rapidly approaches normality as $n$ increases. It also equals a likelihood $F(\mathbf{b}_\lambda; \theta, \lambda) \propto L(\theta; b_\lambda, \lambda) = e^{zb}/(1 + e^z)^n$ from a single binomial observation of $b_\lambda = nr$ successes on $n$ trials when the success probability is $e^z/(1 + e^z)$; thus, our prior-generating experiment is a draw of $b_\lambda$ from the binomial $F(b; \theta, \lambda)$.

The representation $(\mathcal{B}, \mathbf{F}, \mathbf{b}_\lambda)$ will not be unique, reflecting that different experiments may yield the same likelihood function. This is no problem; in fact, alternate representations can help gauge the knowledge claims implicit in the prior $H(\theta; \lambda)$. Translating priors into a likelihood of the form $L(\theta; \mathbf{b}_\lambda, \lambda) = F(\mathbf{b}_\lambda; \theta, \lambda)$ provides a measure of information in $H(\theta; \lambda)$ that can be appreciated in terms of effective sample size (above, the total $n$ in $\mathbf{b}_\lambda$) and other practical features that would produce this information. The exercise thus helps judge whether $H(\theta; \lambda)$ is implausibly informative (Greenland, 2006). It also allows sensitivity analysis based on varying $\mathbf{b}_\lambda$, which may be more intuitive than analyses based on the original parameters in $\lambda$.

Note that conjugacy of $\mathbf{H}$ with the actual data model $\mathbf{G}$ is *not* required: $\mathbf{G}$ and $\mathbf{F}$ may be different distributional families, so that the resulting actual likelihood $L(\gamma, \theta; \mathbf{a})$ and the "prior likelihood" $L(\theta; \mathbf{b}_\lambda, \lambda)$ may be different functional forms related only through $\theta$.

## 2.6 Computation

The penalized loglikelihood $\ln\{L(\gamma, \theta; \mathbf{a}) H(\theta; \lambda)\} = \ln\{L(\gamma, \theta; \mathbf{a}) L(\theta; \mathbf{b}_\lambda, \lambda)\}$ can be summarized in the usual way, with the maximum and negative Hessian (observed penalized information) supplying approximate posterior means and standard deviations (Leonard and Hsu, 1999). When $\ln\{H(\theta; \lambda)\}$ itself decomposes into a sum of "prior likelihood" components, as when $\mathbf{b}_\lambda$ is a vector of independent prior observations, the result is typically a posterior more normal than $L(\gamma, \theta; \mathbf{a})$. This improves the numerical accuracy of posterior tail-area approximations based on the profile-penalized likelihood (Leonard and Hsu, 1999), which can be remarkably close to exact tail areas even with highly skewed distributions (Greenland, 2003b). If $\mathbf{H}$ and $\mathbf{G}$ are conjugate, approximate Bayesian inferences can be obtained simply by appending the prior data $\mathbf{b}_\lambda$ to the actual data $\mathbf{a}$ and entering the augmented data set into ordinary maximum-likelihood software along with appropriate offsets (Greenland, 2003b, 2007a, 2007b, 2009).

Nonetheless, posterior simulation is widely preferred for Bayesian analyses. Markov chain samplers sometimes incur burdens due to autocorrelation and convergence failure, especially when dealing with nonidentified models and improper priors. Under a transparent parameterization with $p(\gamma, \theta|\mathbf{a}) = p(\gamma|\mathbf{a}) p(\theta|\gamma)$ and $G(\mathbf{a}; \gamma, \theta) = G(\mathbf{a}; \gamma)$, we can instead make independent draws from $p(\gamma, \theta|\mathbf{a})$ if we can make independent draws $\gamma^*$ from $p(\gamma|\mathbf{a})$, then draw from $p(\theta|\gamma^*)$. In the application below, $p(\gamma)$ is constant or conjugate with $G(\mathbf{a}; \gamma)$, hence, $p(\gamma|\mathbf{a})$ is conjugate and easy to independently sample when $G(\mathbf{a}; \gamma)$ is a conventional count model. Additionally, with a partial prior $p(\gamma, \theta) = p(\theta)$, or more generally with $p(\gamma, \theta) = p(\gamma) p(\theta)$, drawing from $p(\theta|\gamma)$ reduces to drawing from $p(\theta) = H(\theta; \lambda)$.

# 3. ANALYSES OF A SIDS STUDY

## 3.1 Conventional Analyses

Table 1 presents the relation of maternal antibiotic report during pregnancy $(X)$ to SIDS occurrence $(Y)$ (Kraus, Greenland and Bulterys, 1989). Given the rarity of SIDS, the underlying population risk ratio comparing the exposed to the unexposed $(X = 1$ vs. $X = 0)$ is well approximated by the corresponding odds ratio $OR_{XY}$. Thus, we may take this odds ratio or $\beta = \ln(OR_{XY})$ as the target parameter. The usual maximum-likelihood estimate (MLE) of $OR_{XY}$ is the sample odds ratio $173(663)/134(602) = 1.42$, with standard error for the log $\hat{\beta}$ of $(1/173 + 1/602 + 1/134 + 1/663)^{1/2} = 0.128$ and 95% confidence limits (CL) for $OR_{XY}$ of $\exp\{\ln(1.42) \pm 1.96 \cdot 0.128\} = 1.11, 1.83$. Absent major concerns about bias, such results would commonly be interpreted as providing an inference that $OR_{XY}$ is above 1 but below 2.

Consider next a prior for the odds ratio. At the time of the study only weak speculations could be made. Not even a direction could be asserted: Antibiotics might be associated with elevated risk (marking effects on the fetus of an infection, or via direct effects) or with reduced risk (by reducing presence of infectious agents). Nonetheless, by the time of the SIDS study, US antibiotic prevalence had climbed to 20% over the preceding four decades and yet the SIDS rate remained a fraction of a percent. This high exposure prevalence and the prominence of the outcome effectively ruled out odds ratios on the order of 5 or more because those would have generated notably higher background SIDS risk in earlier studies and surveillance. Thus, one plausible starting prior would have placed $2:1$ odds on $OR_{XY}$ between $\frac{1}{2}$ and 2, and 95% probability on $OR_{XY}$ between $\frac{1}{4}$ and 4. These initial bets follow from a normal$(\mu, \sigma^2)$ prior for $\beta$ that satisfies $\exp(\mu \pm 1.96 \cdot \sigma) = \frac{1}{4}, 4$. Solving, we get $E(\beta) = \mu = 0, \sigma^2 = \frac{1}{2}$, for a penalty of $(\beta - \mu)^2/\sigma^2 = 2\beta^2$, hence subtraction of $\beta^2$ from the loglikelihood.

Let $\mathbf{a}_{XY} = (663, 134, 602, 173)'$ be the vector of counts from Table 1. Without further prior specification, the maximum penalized-likelihood estimate and posterior mode of $\beta$ is 0.341, so $p(\beta|\mathbf{a}_{XY})$ is approximately normal with mean $E(\beta|\mathbf{a}_{XY}) = 0.341$ and standard deviation 0.126. These yield an approximate posterior median for $OR_{XY} = e^{\beta}$ of

$\exp(0.341) = 1.41$ with Wald-type 95% posterior limits of $\exp(0.341 \pm 1.96 \cdot 0.126) = 1.10, 1.80$. These results are barely distinct from the conventional results because the conventional likelihood dominates the prior.

## 3.2 Model Expansion to Accommodate Misclassification

Although the above antibiotic-SIDS prior makes little difference using the conventional likelihood, it makes a profound difference when we expand the likelihood to allow for misclassification. $X$ represents only mother's report of antibiotic use. Let $T$ be the indicator of actual (true) antibiotic use. There is no doubt that mistaken reports $(T \neq X)$ occur. Moreover, recall bias seems likely, with false positives more frequent among cases and false negatives more frequent among controls (more $T < X$ when $Y = 1$, more $T > X$ when $Y = 0$).

Let $A_{txy}$ be the unobserved count variable at $T = t$, $X = x$, $Y = y$, $E_{txy} \equiv E(A_{txy})$ and let a "+" subscript indicate summation over the subscript. $\mathbf{A}_{XY} = (A_{+00}, A_{+10}, A_{+01}, A_{+11})'$ is now the vector of marginal $XY$ count variables with $\mathbf{E}_{XY} \equiv E(\mathbf{A}_{XY}) = (E_{+00}, E_{+10}, E_{+01}, E_{+11})'$. The problem can then be restated as follows: We observe only the $XY$ margin $\mathbf{A}_{XY}$ and get an estimator $A_{+11}A_{+00}/A_{+10}A_{+01}$ of the marginal $XY$ odds ratio $OR_{XY} = E_{+11}E_{+00}/E_{+10}E_{+01}$. But the odds ratio of substantive interest (i.e., the real target parameter $\tau$) is the marginal $TY$ odds ratio $\tau = OR_{TY} = E_{1+1}E_{0+0}/E_{1+0}E_{0+1}$.

With no measurement of $T$, data on $T$ are missing for everyone ($T$ is latent) and $OR_{TY}$ is not identified or even bounded by $\mathbf{A}_{XY}$. To estimate $OR_{TY}$, we need information linking $T$ to $X$ and $Y$, such as prior distributions, subjects with data on $T$ as well as $X$ and $Y$, or both. Examples include information on predicting $T$ from $XY$, that is, information on the predictive values $\pi_{txy} \equiv \Pr(T = t|X = x, Y = y)$. Because $\pi_{0xy} = 1 - \pi_{1xy}$, there are only four distinct classification parameters, which may be taken as $\boldsymbol{\pi}_T = (\pi_{111}, \pi_{110}, \pi_{101}, \pi_{100})'$. Knowing $\boldsymbol{\pi}_T$ would allow us to impute $T$ in the data, as shown in Table 2. Unfortunately, the $XY$ data in Table 1 say nothing about $\boldsymbol{\pi}_T$, that is, $\boldsymbol{\pi}_T$ is not identified by those data. One must impose supplementary constraints to say anything about the target $OR_{TY}$ based on the $XY$ data.

Despite these problems, many epidemiologists anchor inferences for the target parameter tightly around uncorrected estimates. Here, the $OR_{XY}$ estimate

TABLE 1
*Data from case-control study of SIDS (Kraus, Greenland and Bulterys, 1989). X indicates maternal recall of antibiotic use during pregnancy and Y indicates SIDS (Y = 1 for cases, Y = 0 for controls)*

|  | $X = 1$ | $X = 0$ |
|---|---|---|
| $Y = 1$ | 173 | 602 |
| $Y = 0$ | 134 | 663 |

Table 2

*Imputed complete-data table from SIDS study. T indicates actual antibiotic use during pregnancy*

|        | $X = 1$ | | $X = 0$ | |
|--------|---------|---------|---------|---------|
|        | $Y = 1$ | $Y = 0$ | $Y = 1$ | $Y = 0$ |
| $T = 1$ | $173\pi_{111}$ | $134\pi_{110}$ | $602\pi_{101}$ | $663\pi_{100}$ |
| $T = 0$ | $173\pi_{011}$ | $134\pi_{010}$ | $602\pi_{001}$ | $663\pi_{000}$ |
| Totals | 173 | 134 | 602 | 663 |

(1.42, 95% limits 1.11, 1.83) is exactly what one gets for $OR_{TY}$ by assuming $X = T$ (no error in $X$). It is also the answer from a semi-Bayes analysis using a degenerate (single-point-mass) prior for $\boldsymbol{\pi}_T$ that assigns $\Pr(\pi_{11y} = 1) = \Pr(\pi_{10y} = 0) = 1$, an extreme prior which no one holds. In other words, basing inference on the conventional results relies on highly implausible equality constraint; it takes no account of the actual uncertainty or prior information about $\boldsymbol{\pi}_T$, which is vague but at least bounds the $\pi_{1xy}$ away from 0 and 1. The same criticism applies to the conventional Bayesian result (1.41, 95% limits 1.10, 1.80), which are based on the same equality constraint for $\boldsymbol{\pi}_T$.

## 3.3 Loglinear Parameterization

Because nonidentification makes inferences arbitrarily sensitive to the prior, it is essential to consider parameterizations with simple contextual meanings so that sensible priors can be posited. The set of expected counts $E_{txy}$ could be taken as a saturated parameterization for the joint distribution of the $A_{txy}$. One reparameterization that facilitates both prior specification and use of conventional software is

$$
\begin{aligned}
E_{txy}(\boldsymbol{\beta}) = \exp(&\beta_0 + \beta_T t + \beta_X x + \beta_Y y \\
&+ \beta_{TX} tx + \beta_{TY} ty \\
&+ \beta_{XY} xy + \beta_{TXY} txy),
\end{aligned}
$$

(1)

where $\boldsymbol{\beta} = (\beta_0, \beta_X, \beta_Y, \beta_{XY}, \beta_T, \beta_{TX}, \beta_{TY}, \beta_{TXY})'$. Dependence of $E_{txy}$ on $\boldsymbol{\beta}$ will be left implicit below. The $\pi_{1xy}$ follow a saturated logistic model for the regression of $T$ on $X$ and $Y$:

$$
\begin{aligned}
\pi_{1xy} &\equiv \Pr(T = 1 | X = x, Y = y) \\
&= \text{expit}(\beta_T + \beta_{TX} x + \beta_{TY} y + \beta_{TXY} xy),
\end{aligned}
$$

(2)

where $\text{expit}(u) \equiv e^u/(1 + e^u)$. $\boldsymbol{\pi}_T$ is a 1–1 function of the parameter subvector $\boldsymbol{\beta}_T = (\beta_T, \beta_{TX}, \beta_{TY}, \beta_{TXY})'$ of coefficients in the imputation model for the missing $T$ data. In the earlier general notation, $\boldsymbol{\theta} = \boldsymbol{\beta}_T$.

The $TY$ odds ratio when $X = 0$ is $\exp(\beta_{TY})$ and is related to the target $OR_{TY}$ by

$$
\begin{aligned}
R(\boldsymbol{\beta}_X) \\
= OR_{TY} / \exp(\beta_{TY}) \\
= \{1 + \exp(\beta_X + \beta_{TX} + \beta_{XY} + \beta_{TXY})\} \\
\cdot \{1 + \exp(\beta_X)\} \\
/ \{1 + \exp(\beta_X + \beta_{TX})\}\{1 + \exp(\beta_X + \beta_{XY})\},
\end{aligned}
$$

where $\boldsymbol{\beta}_X = (\beta_X, \beta_{TX}, \beta_{XY}, \beta_{TXY})'$. The latter expression is a factor for a problem in which $X$ is a confounder rather than a measurement of $T$ (Yanagawa, 1984); it can also be used to represent selection bias (see below). Here it is useful for deriving the prior for $\beta_{TY}$ from priors or constraints on $\boldsymbol{\beta}_X$ and $OR_{TY}$. For example, ascertainment of $X$ before $Y$ occurs may lead to $X \perp\!\!\!\perp Y | T$ (nondifferential misclassification), which is equivalent to $\beta_{XY} = \beta_{TXY} = 0$; in that case $R(\boldsymbol{\beta}_X) = 1$ and $\exp(\beta_{TY}) = OR_{TY}$, so that the priors on $\exp(\beta_{TY})$ and $OR_{TY}$ must be identical. Nondifferentiality can be relaxed by using priors centered at zero for $\beta_{XY}$ and $\beta_{TXY}$.

## 3.4 Transparent Reparameterization

The full model involves 8 parameters for the 4 observations, and no component of $\boldsymbol{\beta}$ is identified without some constraint. We can, however, reparameterize the saturated model into an identified parameter $\mathbf{E}_{XY}$ and the nonidentified $\boldsymbol{\pi}_T$ or (equivalently) $\boldsymbol{\beta}_T$,

$$
\begin{aligned}
E_{txy} &= E_{+xy} \pi_{txy} \\
&= E_{+xy} \, \text{expit}(\beta_T + \beta_{TX} tx + \beta_{TY} ty \\
&\qquad\qquad + \beta_{TXY} txy).
\end{aligned}
$$

(3)

In the general notation we have $\mathbf{a} = \mathbf{a}_{XY}$, $\boldsymbol{\gamma} = \mathbf{E}_{XY}$, $\boldsymbol{\theta} = \boldsymbol{\beta}_T$, and $G(\mathbf{a}; \boldsymbol{\gamma}, \boldsymbol{\theta}) = \Pr(\mathbf{A}_{XY} = \mathbf{a}_{XY} | \mathbf{E}_{XY}, \boldsymbol{\beta}_T) = \Pr(\mathbf{A}_{XY} = \mathbf{a}_{XY} | \mathbf{E}_{XY})$. The likelihood depends solely on the identified parameter $\mathbf{E}_{XY}$ (i.e., for any constant $\mathbf{c}$, $\mathbf{E}_{XY} = \mathbf{c}$ defines a level set of the likelihood surface). It follows that there is no updating of $p(\boldsymbol{\beta}_T | \mathbf{E}_{XY})$, that is, $p(\boldsymbol{\beta}_T | \mathbf{E}_{XY}, \mathbf{A}_{XY}) = p(\boldsymbol{\beta}_T | \mathbf{E}_{XY})$, hence, $p(\boldsymbol{\beta}_T, \mathbf{E}_{XY} | \mathbf{A}_{XY}) = p(\boldsymbol{\beta}_T | \mathbf{E}_{XY}) p(\mathbf{E}_{XY} | \mathbf{A}_{XY})$.

Because $E_{t+y} = E_{+1y} \pi_{t1y} + E_{+0y} \pi_{t0y}$, the target parameter $OR_{TY} = E_{1+1} E_{0+0} / E_{1+0} E_{0+1}$ is a mixture of identified parameters $E_{+xy}$ and nonidentified $\pi_{txy}$. Thus, $OR_{TY}$ may be updated both through $p(\mathbf{E}_{XY} | \mathbf{A}_{XY})$ and $p(\boldsymbol{\beta}_T | \mathbf{E}_{XY})$; but with $p(\boldsymbol{\beta}_T | \mathbf{E}_{XY}) = p(\boldsymbol{\beta}_T)$, as here, the update will involve only $p(\mathbf{E}_{XY} | \mathbf{A}_{XY})$.

TABLE 3
*Parameters of normal priors for coefficients in the logistic regression of $T$ on $X$ and $Y$ (prior for $\beta_{TX} + \beta_{TXY}$ induced by priors for $\beta_{TX}$ and $\beta_{TXY}$)*

| | Mean | Variance | 95% prior limits for | $n = 2b = 4/\text{variance}^*$ |
|---|---|---|---|---|
| $\beta_T$ | $\text{logit}(0.1)$ | 0.16 | $\text{expit}(\beta_T)$: 0.05, 0.20 | 25 |
| $\beta_{TX}$ | $\ln(13.5)$ | 0.25 | $\exp(\beta_{TX})$: 5, 36 | 16 |
| $\beta_{TY}$ | 0 | 0.50 | $\exp(\beta_{TY})$: $\frac{1}{4}$, 4 | 8 |
| $\beta_{TXY}$ | 0 | 0.125 | $\exp(\beta_{TXY})$: $\frac{1}{2}$, 2 | 32 |
| $\beta_{TX} + \beta_{TXY}$ | $\ln(13.5)$ | 0.375 | $\exp(\beta_{TX} + \beta_{TXY})$: 4.1, 45 | (not used) |

*Number of binomial trials needed to make asymptotic variance estimate of $\text{logit}(B/n)$ equal to prior variance when the number of successes $B$ is $b = n/2$; used for penalized estimation.

## 3.5 An Initial Prior Specification

As with specifications for regression models, no prior distribution could be claimed "correct." Nonetheless, some specifications are plausible and others are not in light of background information. For the nonidentified $T$-predictive parameter $\boldsymbol{\beta}_T$, consider first $\Pr(T = 1|X = 0)$, the probability among noncases that a "test negative" ($X = 0$) is erroneous. Because of SIDS rarity we have $\Pr(T = 1|X = 0) \approx \Pr(T = 1|X = 0, Y = 0) = \text{expit}(\beta_T) = \pi_{100}$. Antibiotic prevalence $\Pr(T = 1)$ in unselected pregnancies was expected to be well below 50%, hence, we should expect $\pi_{100}$ to be small but nonzero to reflect false negatives. These considerations suggest that plausible distributions for $\text{expit}(\beta_T)$ include some placing 95% probability between 0.05 and 0.20.

Next, let $\varphi_{xty} = \Pr(X = x|T = t, Y = y)$. Then $\varphi_{1ty} = \text{expit}(\beta_X + \beta_{TX}t + \beta_{XY}y + \beta_{TXY}ty)$ and the $Y$-specific receiver-operating characteristic (ROC) odds ratios (true-positive odds/false-positive odds) are

$$OR_{TX}(y) = (\varphi_{11y}/\varphi_{01y})/(\varphi_{10y}/\varphi_{00y})$$
$$= (\pi_{11y}/\pi_{10y})/(\pi_{01y}/\pi_{00y})$$
$$= \exp(\beta_{TX} + \beta_{TXY}y).$$

If $X$ is pure noise, a $p$-coin flip, then $\varphi_{11y} = \varphi_{10y} = p$, $\beta_{TX} = \beta_{TXY} = 0$ and $OR_{TX}(y) = 1$. Background literature (e.g., Werler et al., 1989) suggests $X$ is nowhere near this bad. Plausible values for $\varphi_{110}$ include 0.6 and 0.8, and for $\varphi_{100}$ include 0.1 and 0.2, hence, plausible values for $OR_{TX}(0) = \exp(\beta_{TX})$ include $(0.6/0.4)/(0.2/0.8) = 6$, $(0.6/0.4)/(0.1/0.9) = 13.5$, $(0.8/0.2)/(0.2/0.8) = 16$, $(0.8/0.2)/(0.1/0.9) = 36$, suggesting that plausible distributions for $\exp(\beta_{TX})$ include some with at least 95% probability between 5 and 40.

The greater uncertainty about $OR_{TX}(1) = \exp(\beta_{TX} + \beta_{TXY})$, the ROC odds ratio among cases, is captured by $OR_{TX}(1)/OR_{TX}(0) = \exp(\beta_{TXY})$, which exceeds 1 if cases have more accurate recall on balance than noncases and is under 1 if vice-versa. A common assumption is that the misclassification is nondifferential, that is, that $X$ and $Y$ are independent given $T$, or equivalently, equal sensitivity and specificity across $Y$. Because $\beta_{XY}$ and $\beta_{XY} + \beta_{TXY}$ are the $XY$ log odds ratios in the $T = 0$ and $T = 1$ strata, under nondifferentiality we have $\beta_{XY} = \beta_{TXY} = 0$, making $\exp(\beta_{TXY}) = 1$, $R(\boldsymbol{\beta}_X) = 1$, and hence, $OR_{TY} = \exp(\beta_{TY})$ (Greenland, 2003c).

Self-report $X$ could be affected by the outcome $Y$, hence, nondifferentiality is not a justifiable assumption. Nonetheless, it is plausible that the impact of $Y$ on $X$ is limited. Furthermore, it is difficult to predict which of cases or noncases would have a higher ROC odds ratio: Cases may have improved recall of true exposure ($\varphi_{111} > \varphi_{110}$) but also more false exposure recall ($\varphi_{011} > \varphi_{010}$), which have opposing effects on $\exp(\beta_{TXY})$. In line with these considerations, placing 95% probability on $\exp(\beta_{TXY})$ between $\frac{1}{2}$ and 2 provides a modest expansion for the distribution of $OR_{TX}(1)$ beyond that of $OR_{TX}(0)$.

These considerations are also relevant to $\beta_{TY}$. If the departures from nondifferentiality are limited, the departures of $\beta_{XY}$ and $\beta_{TXY}$ from zero are small, which in turn implies that $R(\boldsymbol{\beta}_X)$ is small and, hence, $\exp(\beta_{TY})$ is close to $OR_{TY}$ (Greenland, 2003c). These results suggest using a prior for $\beta_{TY}$ similar to that for $OR_{TY}$, for example, a lognormal prior with 95% probability between $\frac{1}{4}$ and 4.

Table 3 presents a set of normal priors that are consistent with the preceding considerations, along with the implied density for $\beta_{TX} + \beta_{TXY}$. The corresponding joint prior density is independent-normal with mean $\boldsymbol{\mu}_T \equiv (\mu_T, \mu_{TX}, \mu_{TY}, \mu_{TXY})' = (\text{logit}(0.1),$ $\ln(13.5), 0, 0)'$ and covariance matrix with diagonal $\boldsymbol{\nu}_T \equiv (\nu_T, \nu_{TX}, \nu_{TY}, \nu_{TXY})' = (0.16, 0.25, 0.50,$ $0.125)'$. In the general notation with $\boldsymbol{\theta} = \boldsymbol{\beta}_T$ and $\boldsymbol{\lambda} = [\boldsymbol{\mu}_T, \boldsymbol{\nu}_T]$, the joint prior density $H(\boldsymbol{\beta}_T; \boldsymbol{\lambda})$ corresponds to the penalty $-2\ln\{H(\boldsymbol{\beta}_T; \boldsymbol{\lambda})\} = \Sigma_i (\beta_i - \mu_i)^2/\nu_i$ where $i = T, TX, TY, TXY$.

As one gauge of prior information, Table 3 shows the number $n_i = 4/\nu_i$ of Bernoulli($\frac{1}{2}$) trials with $B$ "successes" that would make the approximate sampling variance $1/\{n_i(\frac{1}{2})(\frac{1}{2})\} = 4/n_i$ of logit($B/n_i$) equal to the prior variance $\nu_i$. One can penalize $\beta_i$ with ordinary maximum-likelihood logistic-regression software by entering a data record with $b_i = 2/\nu_i$ "successes" out of $n_i = 4/\nu_i$ trials, zero for all covariates except $i$ (for which it is 1) and an offset $-\mu_i$ (Greenland, 2007a). The result is a binomial likelihood contribution

$$L_i \equiv L(\beta_i; b_i) \propto \text{expit}(\beta_i - \mu_i)^{b_i} \text{expit}(-\beta_i + \mu_i)^{b_i}$$
$$= \exp(\beta_i - \mu_i)^{b_i}/\{1 + \exp(\beta_i - \mu_i)\}^{2b_i},$$

which is close to normal for $n_i \geq 8$, becoming heavier-tailed for smaller $n_i$. With $\mathbf{b}_{\boldsymbol{\lambda}} = (b_T, b_{TX}, b_{TY}, b_{TXY})' = 2/\boldsymbol{\nu}_T$, we obtain $H(\boldsymbol{\beta}_T; \boldsymbol{\lambda}) = F(\mathbf{b}_{\boldsymbol{\lambda}}; \boldsymbol{\beta}_T, \boldsymbol{\mu}_T) \propto L(\boldsymbol{\beta}_T; \mathbf{b}_{\boldsymbol{\lambda}}, \boldsymbol{\mu}_T) = \Pi_i L_i$.

As another gauge of prior information, $b_i$ may also be interpreted as the number of cases one would have to observe in each arm of a randomized trial of a treatment and a rare outcome with allocation ratio $\exp(-\mu_i)$ to obtain an approximate variance of $\nu_i$ for the log odds-ratio estimate (Greenland, 2006). More generally, as mentioned earlier, one can derive the $b_i$, $n_i$, and offset or allocation needed to produce a likelihood that is exactly proportional to a generalized log-$F$ density with $2b_i$ and $2(n_i - b_i)$ degrees of freedom; this extension allows skewness or heavy tails for the prior (Greenland, 2003b, 2007b).

### 3.6 Penalized-Likelihood and Posterior-Sampling Analyses

Using the loglinear parameterization and prior likelihoods $L_i$, the penalized likelihood is $L(\boldsymbol{\beta}|\mathbf{a}_{XY}, \boldsymbol{\lambda}) = L(\boldsymbol{\beta}; \mathbf{a}_{XY})H(\boldsymbol{\beta}_T; \boldsymbol{\lambda}) = L(\boldsymbol{\beta}; \mathbf{a}_{XY})L(\boldsymbol{\beta}_T; \mathbf{b}_{\boldsymbol{\lambda}}, \boldsymbol{\mu}_T)$, in which $L(\boldsymbol{\beta}; \mathbf{a}_{XY})$ derives from actual-data records with $T$ missing, and $L(\boldsymbol{\beta}_T; \mathbf{b}_{\boldsymbol{\lambda}}, \boldsymbol{\mu}_T)$ derives from hypothetical complete-data records. Analysis can then proceed using standard likelihood methods for missing data (McLachlan and Krishnan, 1997; Little and Rubin, 2002). Alternatively, using the transparent parameterization, we obtain independent draws from

the exact marginal posterior $p(OR_{TY}|\mathbf{a}_{XY})$ as follows: (1) draw $\mathbf{E}_{XY}^*$ from $p(\mathbf{E}_{XY}|\mathbf{a}_{XY})$; (2) draw $\boldsymbol{\beta}_T^*$ from $p(\boldsymbol{\beta}_T|\mathbf{E}_{XY}^*)$; (3) compute $\boldsymbol{\pi}_T^*$ from $\boldsymbol{\beta}_T^*$, $E_{t+y}^* = E_{+1y}^*\pi_{t1y}^* + E_{+0y}^*\pi_{t0y}^*$, and $OR_{TY}^* = E_{+1}^*E_{+0+}^*/E_{+0}^*E_{0+1}^*$. With a noninformative prior for $\mathbf{E}_{XY}$ and $p(\boldsymbol{\beta}_T|\mathbf{E}_{XY}) = p(\boldsymbol{\beta}_T)$, as here, step (2) reduces to drawing $\boldsymbol{\beta}_T^*$ from $p(\boldsymbol{\beta}_T)$ and the resulting sampler is approximated by Monte Carlo Sensitivity Analysis (MCSA) in which bootstrap draws $\mathbf{a}_{XY}^*$ from $\mathbf{a}_{XY}$ replace $\mathbf{E}_{XY}^*$ (Greenland, 2005a).

Using the partial prior $p(\boldsymbol{\beta}_T)$ in Table 3 with $A_{txy}$ either Poisson or multinomial conditional on the $Y$ margin $(a_{++1}, a_{++0})$, the penalized-likelihood estimate for $OR_{TY}$ is 1.19, with Wald 95% limits of 0.41, 3.43. Both exact posterior sampling and MCSA with 250,000 draws yield a median for $OR_{TY}$ of 1.19 with 2.5th and 97.5th percentiles of 0.37, 3.42, not much different in practical terms from the 95% prior limits (0.25, 4). The posterior variance of $OR_{TY}$ is more sensitive to the prior variance $\nu_{TY}$ for $\beta_{TY}$ than to the other prior variances. Upon increasing $\nu_{TY}$ to make the prior 95% limits for $\exp(\beta_{TY})$ equal to 0.125, 8, the 2.5th and 97.5th sampling percentiles for $OR_{TY}$ become 0.20, 6.1, again not much different from the prior in practical terms. A common response to this variance sensitivity would be to set a hyperprior on the prior variance $\nu_{TY}$; that would, however, obscure both the contextual meaning of the prior and the extreme sensitivity of the results to $\nu_{TY}$.

In examining these results, there are several ways to contrast the contribution of $p(\boldsymbol{\beta}_T|\mathbf{E}_{XY})$, which represents uncertainty about $\boldsymbol{\pi}_T$, against the contribution of $p(\mathbf{E}_{XY}|\mathbf{a}_{XY})$, which represents uncertainty about $\mathbf{E}_{XY}$. One natural way to gauge the contribution of $p(\boldsymbol{\beta}_T|\mathbf{E}_{XY})$ is to contrast posterior intervals, such as the 95% posterior sampling interval (0.37, 3.42), against analogous intervals that assume no uncertainty about $\boldsymbol{\pi}_T$, such as the conventional 95% interval (1.11, 1.83). Another way is to take the ratio of the estimated sampling variance of $\ln(OR_{XY})$ to the posterior variance of $\ln(OR_{TY})$, which here is 5.6%. Either way, the results show that the precision of the conventional frequentist and Bayesian results is due entirely to the equality constraints on the predictive values $\pi_{txy}$, that is, $p(\pi_{11y} = 1) = p(\pi_{00y} = 1) = 1$. This is unsurprising insofar as $OR_{TY}$ is not identified by the expanded likelihood; hence, the data add little information about $OR_{TY}$ beyond that in the prior.

### 3.7 Dependent Parameterizations

The general ideas discussed so far apply to arbitrary parameterizations of the data model $G(\mathbf{a}; \boldsymbol{\gamma}, \boldsymbol{\theta})$, including those in which $\boldsymbol{\theta}$ is partially identified through dependence on $\boldsymbol{\gamma}$. In the misclassification problem an example occurs when the bias parameter $\boldsymbol{\theta}$ is taken to be $\boldsymbol{\beta}_X = (\beta_X, \beta_{TX}, \beta_{XY}, \beta_{TXY})'$ rather than $\boldsymbol{\beta}_T = (\beta_T, \beta_{TX}, \beta_{TY}, \beta_{TXY})'$. Specification of $\boldsymbol{\theta} = \boldsymbol{\beta}_X$ and its prior follows naturally when the initial priors are for the true and false positive probabilities (the $\varphi_{1ty}$), because these probabilities are functions solely of $\boldsymbol{\beta}_X$: $\varphi_{1ty} = \mathrm{expit}(\beta_X + \beta_{TX}t + \beta_{XY}y + \beta_{TXY}ty)$. However, the identified expectations in $\boldsymbol{\gamma} = \mathbf{E}_{XY}$ imply bounds on the $\varphi_{1ty}$; hence, $\mathbf{E}_{XY}$ constrains $\boldsymbol{\beta}_X$ and the data $\mathbf{A}_{XY}$ identify these constraints. In other words, unlike $\mathbf{E}_{XY}$ and $\boldsymbol{\beta}_T$, $\mathbf{E}_{XY}$ and $\boldsymbol{\beta}_X$ are variation dependent, which can viewed as a logical prior dependence.

Such dependence can be handled by general fitting methods (Joseph, Gyorkos and Coupal, 1995; Gustafson, Le and Saskin, 2001; Gustafson, 2003), but invalidates simplified posterior computations like MCSA that assume no updating of the bias parameters. And although a proper prior on $\boldsymbol{\beta}_X$ will identify the target parameter $OR_{TY}$, it will lead to an improper posterior for the full parameter $\boldsymbol{\beta}$ if no further prior specification is made (see Gelfand and Sahu, 1999, for more general results along these lines). Thus, if a dependent parameterization is preferred (say, for ease of prior specification), one way to proceed is to penalize as necessary to ensure identification of the full parameter vector and employ fitting methods that do not assume prior independencies. As illustrated in Section 3.5, however, one could instead retain the transparent parameterization (and the simplifications its use entails) by incorporating prior information on the $\varphi_{1ty}$ into the prior specification process for $\boldsymbol{\beta}_T$.

### 3.8 How Conditional Should the Probability Model Be?

Log-linear analysis of case-control counts was introduced over 30 years ago (Bishop, Fienberg and Holland, 1975) but appears to have been forgotten in favor of logistic regression with $Y$ as the outcome, which developed in the same era. The history is unsurprising: Unlike logistic regression, the usual log-linear approach requires categorization of continuous covariates and is quite limited in the number of covariates it can handle. Furthermore, in case-control studies (in which the $Y$ total is constrained by design), penalization does not affect the consistency of odds-ratio estimates from logistic regression with $Y$ as the outcome (Greenland, 2003b, Section 3). Nonetheless, the log-linear approach provides a model of the joint distribution of observed and latent variables in a single regression, which can greatly simplify bias analysis, hence its resurrection here.

Most literature on tabular data adheres to models that condition on the total or a margin of the tabular data, even when those quantities are not fixed by design. This practice creates no issue for odds-ratio analyses: One obtains identical likelihood-based inferences on odds ratios from multinomial or binomial (conditional) and Poisson (unconditional) sampling models that include fixed margins as unconstrained log-linear effects (Bishop, Fienberg and Holland, 1975, Sections 3.5 and 13.4.4). Nonetheless, these design effects imply that no prior should be placed on parameters that are functions of sample size or sampling ratios. For example, in a case-control study the log-linear intercept and disease coefficient, $\beta_0$ in $\beta_Y$ model (1), are functions of the design and so should receive no prior. These considerations are a further reason for adopting the partial-prior (semi-Bayes) analysis used here.

## 4. EXTENSIONS AND GENERALIZATIONS

The above formulation has straightforward extensions to other biases and more general data models. These are sketched briefly here.

### 4.1 Validation and Alternative Measurements

Adding a plausible measurement model shows that the SIDS data offer far less information about the target $OR_{TY}$ than the conventional analysis makes it seem. Sharper inference about $OR_{TY}$ requires sharper information about the $TXY$ distribution. Short of measuring $T$ directly on everyone in the study, such information might come from an alternate measurement $W$ of $T$ on a sample from the source population of the study, along with $X$ and $Y$. If this measurement is error-free ($W = T$) or assumed so, the alternate measurements are called "validation data" (Carroll et al., 2006) and yield actual-data records with $T$ present.

Unfortunately, validation data are often unavailable, impractical to obtain in a timely manner or inadequate in quantity. Then too, they suffer their own errors and biases. Subjects do not randomly refuse further study, and alternate measurements have errors ($W \neq T$). Thus, regardless of the data available, an accurate uncertainty assessment requires an expanded model to link the observations ($TXY$ and the partially observed $W$) to unobserved variables ($T$ and missing $W$). The identification problem is not removed, rather $W$ is added to certain records, whose informativeness depends entirely on the priors relating them to the target variables.

TABLE 4
*SIDS data separated into strata with prescription examined in medical record ("validated," assuming $W = T$) and remainder: $W = 1$ if record shows prescription, $W = 0$ if not*

| | $Y = 1$ | | $Y = 0$ | |
|---|---|---|---|---|
| | $X = 1$ | $X = 0$ | $X = 1$ | $X = 0$ |
| Medical-record data on $W$: | | | | |
| $W = 1$ | 29 | 17 | 21 | 16 |
| $W = 0$ | 22 | 143 | 12 | 168 |
| Totals | 51 | 160 | 33 | 184 |
| $\hat{\pi}_{1xy}$ | 0.569 | 0.106 | 0.636 | 0.087 |
| No $W$ data: | | | | |
| $W$ missing | 122 | 442 | 101 | 479 |
| Imputed counts (for $W = 1$, $\hat{\pi}_{1xy}$ times $W$-missing count; for $W = 0$, $\hat{\pi}_{0xy}$ times $W$-missing count): | | | | |
| $W = 1$ | 73.2 | 44.2 | 60.6 | 47.9 |
| $W = 0$ | 48.8 | 397.8 | 40.4 | 431.1 |

In the SIDS example, a pseudo-random sample of medical records was used to check maternal responses in the subsample (Drews, Kraus and Greenland, 1990). $W$ is the record indicator for antibiotic prescription. Table 4 shows the data from Table 1 separated into $W$-known (alternate or complete-data, with $W = 1$ or 0) and $W$-unknown (incomplete-data) strata. If $W = T$, the resulting data provide a likelihood for the $T|XY$ parameter vector $\boldsymbol{\pi}_T$ or $\boldsymbol{\beta}_T$. Because the $TXY$ model is saturated, maximum likelihood simplifies to using the MLEs $\hat{\pi}_{txy}$ from Table 4 in place of the $\pi_{txy}$ in Table 2 to impute $T$ where it is missing, followed by collapsing over $X$ (Lyles, 2002). The resulting marginal $TY$ odds ratio $\hat{OR}_{TY}$ is the MLE of $OR_{TY}$. For unsaturated models, $\hat{OR}_{TY}$ has no closed form but may be replaced by any reasonably efficient closed-form estimator (Greenland, 2007c); otherwise a full likelihood method may be used (Espeland and Hui, 1987; Little and Rubin, 2002; Carroll et al. 2006).

Assuming $W = T$, $\hat{OR}_{TY} = 1.21$ with Wald 95% confidence limits 0.79, 1.87. Adding the partial prior in Table 3, an approximate posterior median for $OR_{TY}$ is 1.20 with 95% Wald limits of 0.81, 1.77, close to the results without the prior. This result shows that the prior is considerably less informative than the record data when we assume $W = T$. But, as with $X = T$, the constraint $W = T$ is unjustified: First, the records only show prescription, not compliance (hence, we should expect for some women $T < W$); second, the records must have some errors due to miscellaneous oversights (e.g., miscoding).

Even if we assume that oversights are negligible, the effect of $W$ is a prescribing (intention-to-treat) effect, and thus (due to noncompliance) is likely biased for the biologic effect of $T$. As before, if we lack $T$ for samples of cases and controls, identification of $OR_{TY}$ depends entirely on priors for $\pi_{1wxy} \equiv \Pr(T = 1|W = w, X = x, Y = y)$. Frequentist analyses arise from sets of equality constraints (point priors) that identify the parameter of interest. $W = T$ is sufficient by itself but implausible, so less strict constraints such as $\text{cov}(W, X|T) = 0$ may be introduced along with other conditions as needed for identification (Hui and Walter, 1980; Carroll et al., 2006; Messer and Natarajan, 2008).

Additional data sources or variables may also provide partial identification (Johnson, Gastwirth and Pearson, 2001; Small and Rosenbaum, 2009). But without such information, results depend on the $\pi_{1wxy}$ priors in an unlimited fashion: With noninformative priors for the $\pi_{1wxy}$, we obtain a noninformative posterior for $OR_{TY}$. If these priors are only vaguely informative, as those above, the posterior distribution for $OR_{TY}$ will be very dispersed.

I omit an extended ($TWXY$) analysis because it would merely illustrate again how posterior concentration is purchased by using extremely informative priors even when alternate measurements are made. Such priors may sometimes be plausible. Nonetheless, in many situations the true exposure history can never be known without considerable potential for systematic error (e.g., lifetime occupational exposures, environmental exposures and nutrient intakes). In these situations, equality constraints need to be recognized as priors, because failure to do so risks overconfident inferences.

These remarks should not be taken as discouraging collection of additional predictors of the true exposure $T$, since such data provide an empirical basis for addressing measurement issues. The present discussion merely cautions against overlooking the nonidentified elements in any model for their use.

### 4.2 Unmeasured Confounders

Consider a setting in which $X$ rather than $T$ is the exposure variable of interest and $T$ is an unmeasured confounder of the effect of $X$ on $Y$. The target effect is now that of $X$ on $Y$; nonetheless, the regression models used for misclassification can be applied unchanged. This effect may be parameterized by the pair of $T$-conditional $X$–$Y$ odds ratios $\exp(\beta_{XY})$ and $\exp(\beta_{XY} + \beta_{TXY})$. It is usually assumed that these odds ratios are equal ($\beta_{TXY} = 0$), leaving $\exp(\beta_{XY})$ as the target; although this equality is another unjustified point prior, it may incur only minor bias in estimating summary effects (Greenland and Maldonado, 1994). With the assumption, the $T$-adjusted odds ratio $\exp(\beta_{XY})$ is related to the unadjusted odds ratio $OR_{XY}$ by

$$R(\boldsymbol{\beta}_T) = OR_{XY} / \exp(\beta_{XY})$$
$$= \{1 + \exp(\beta_T + \beta_{TX} + \beta_{TY})\}\{1 + \exp(\beta_T)\}$$
$$/ \{1 + \exp(\beta_T + \beta_{TX})\}\{1 + \exp(\beta_T + \beta_{TY})\}$$

(Yanagawa, 1984). Without data on $T$, $\boldsymbol{\beta}_T$ and hence $R(\boldsymbol{\beta}_T)$ are not identified. Thus, assuming $p(\boldsymbol{\beta}_T | \mathbf{E}_X) = p(\boldsymbol{\beta}_T) = H(\boldsymbol{\beta}_T; \boldsymbol{\lambda})$, to draw $\exp(\beta_{XY}^*)$ from $p\{\exp(\beta_{XY}) | \mathbf{a}_{XY}\}$, we draw $\mathbf{E}_{XY}^*$ from $p(E_{XY} | \mathbf{a}_{XY})$, compute $OR_{XY}^* = E_{+11}^* E_{+00}^* / E_{+10}^* E_{+01}^*$, draw $\boldsymbol{\beta}_T^*$ from $H(\boldsymbol{\beta}_T; \boldsymbol{\lambda})$, and compute $\exp(\beta_{XY}^*) = OR_{XY}^* / R(\boldsymbol{\beta}_T^*)$. Assuming $p(\boldsymbol{\beta}_T, \mathbf{E}_{XY}) = p(\boldsymbol{\beta}_T)$, MCSA uses bootstrap draws $\mathbf{a}_{XY}^*$ from $\mathbf{a}_{XY}$ in place of $\mathbf{E}_{XY}^*$ (Greenland, 2003a).

In two-stage (two-phase) studies, $T$ is measured on subsamples of subjects randomly selected within $X$–$Y$ levels (White, 1982; Walker, 1982). This design is formally identical to validation subsampling; the resulting complete records may be entered into the analysis as described earlier.

### 4.3 Selection Bias

Consider again a setting in which $T$ is the exposure variable of interest. Let $X = 1 - S$ where $S$ is the selection indicator, so only subjects with $X = 0$ are observed. The models and target parameter $OR_{TY}$ used for misclassification are unchanged, but now observed records are complete (include $T, X, Y$) and they are confined to the $X = 0$ stratum. The observations are $\mathbf{a}_0 = (a_{101}, a_{100}, a_{001}, a_{000})'$.

With no data at $X = 1$, the log-linear parameterization is transparent with identified component $\boldsymbol{\gamma} = (\beta_0, \beta_T, \beta_Y, \beta_{TY})'$ and nonidentified $X | TY$ component $\boldsymbol{\theta} = \boldsymbol{\beta}_X = (\beta_X, \beta_{TX}, \beta_{XY}, \beta_{TXY})'$. The $TY$ odds-ratio parameter in the $X = 0$ stratum is $\exp(\beta_{TY}) = E_{101} E_{000} / E_{100} E_{001}$, and is related to the target by $OR_{TY} = \exp(\beta_{TY}) R(\boldsymbol{\beta}_X)$. Thus, assuming $p(\boldsymbol{\beta}_X | \mathbf{E}_0) = p(\boldsymbol{\beta}_X) = H(\boldsymbol{\beta}_X; \boldsymbol{\lambda})$, to draw $OR_{TY}^*$ from $p\{OR_{TY} | \mathbf{a}_0\}$, we draw $\mathbf{E}_0^*$ from $p(E_0 | \mathbf{a}_0)$, compute $\exp(\beta_{TY}^*) = E_{101}^* E_{000}^* / E_{100}^* E_{001}^*$, draw $\boldsymbol{\beta}_X^*$ from $H(\boldsymbol{\beta}_X; \boldsymbol{\lambda})$, and compute $OR_{TY}^* = \exp(\beta_{TY}^*) R(\boldsymbol{\beta}_X^*)$. Assuming $p(\boldsymbol{\beta}_X, \mathbf{E}_{XY}) = p(\boldsymbol{\beta}_X)$, MCSA uses bootstrap draws $\mathbf{a}_0^*$ from $\mathbf{a}_0$ in place of $\mathbf{E}_0^*$. If, however, selection is modeled as a Poisson process with $TY$-dependent sampling rate $\exp(\beta_S + \beta_{ST} t + \beta_{SY} y + \beta_{STY} ty)$, as in "density" (risk-set) sampling, $R(\boldsymbol{\beta}_X)$ simplifies to $\exp(-\beta_{STY})$, hence, one need only specify and sample from $p(\beta_{STY})$ (Greenland, 2003a).

Occasionally, information on nonresponders (subjects with $X = 1$) becomes available. Such information may arise from general records or from call-back surveys of nonrespondents. Nonetheless, respondents in call-back surveys are unlikely to be a random sample of all the original nonrespondents, hence, further parameters will be needed to relate survey exclusion to $T$, $X$ and $Y$.

### 4.4 Multiple Biases and Multiple Variables

The above approach supplements the observed variables $\mathbf{Z}$ (representing available measurements) with wholly latent variables $\mathbf{T}$ (representing unobserved target variables and unmeasured confounders). It then formulates an identified observable model $P(\mathbf{z} | \boldsymbol{\gamma})$, a selection-rate model $S(\mathbf{t}, \mathbf{z}; \boldsymbol{\beta}_S)$, an imputation model $P(\mathbf{t} | \mathbf{z}; \beta_T)$ for $\mathbf{T}$ and a plausible prior $H(\boldsymbol{\theta}; \boldsymbol{\lambda})$ for the nonidentified $\boldsymbol{\theta} = (\boldsymbol{\beta}_S, \beta_T)$. Inference to population quantities involving $\mathbf{T}$ can then be based on $p(\mathbf{t}, \mathbf{z}) \propto P(\mathbf{t} | \mathbf{z}; \beta_T) P(\mathbf{z} | \boldsymbol{\gamma}) / S(\mathbf{t}, \mathbf{z}; \boldsymbol{\beta}_S)$. For discrete data, we may replace $P(\mathbf{z} | \boldsymbol{\gamma})$ with $\mathbf{E}_z(\boldsymbol{\gamma})$, the expected data count at $\mathbf{Z} = \mathbf{z}$. With $p(\boldsymbol{\gamma}, \boldsymbol{\theta}) = p(\boldsymbol{\gamma}) p(\boldsymbol{\theta})$, posterior sampling reduces to sampling from $p(\boldsymbol{\gamma} | \mathbf{z}) \times H(\boldsymbol{\theta}; \boldsymbol{\lambda})$; with an improper prior $p(\boldsymbol{\gamma}, \boldsymbol{\theta}) = p(\boldsymbol{\theta}) =$ $H(\boldsymbol{\theta}; \boldsymbol{\lambda})$, we may replace $P(\mathbf{z}|\boldsymbol{\gamma})$ by its bootstrap estimate. Addition of identifying data on $T$ is handled in the more usual Bayesian framework.

The general approach models the joint distribution of all variables in the problem, including several wholly latent variables; thus, the number of parameters can become huge. Effective degrees of freedom can be reduced via hierarchical modeling of the parameters (Greenland, 2003a, 2005a); for example, Greenland and Kheifets (2006) analyzed 60 observed counts with hierarchical models that included 135 first-stage (data-level) bias parameters generated from second-stage linear models. The profusion of parameters reflects a reality of observational research hidden by conventional analyses, which implicitly set most parameters to zero. Nonetheless, uncertainty can often be addressed adequately by rather simple analyses of one or two biases; in the example, those analyses quickly reveal that the data cannot sustain any accurate inference about the target parameter given uncertainties about the bias sources.

### 4.5 Semi-parametric Modeling

Semi-parametric methods have been extended to incorporate nonidentified confounding and selection biases when these biases reduce to simple multiplicative or additive forms (e.g., Brumback et al., 2004; Robins, Rotnitzky and Scharfstein, 2000; Scharfstein, Rotnitsky and Robins, 1999). Adjustment factors in these extensions correspond to model-based factors such as $R(\boldsymbol{\beta}_X)$, but lack the finer parametric structure of the latter. As illustrated above, the parameters within these factors can serve in prior specification from background information. This role is important insofar as prior specification is the hardest task in bias modeling, especially because noninformative and other reference priors are not serious options for nonidentified parameters.

Note that semi-parametric robustness is achieved by only partially specifying the distribution of *observables*, and thus does not extend to specification of nonidentified model components. Nonetheless, the approach illustrated here can be used to extend semi-parametric models by penalizing the partial- or pseudo-loglikelihood, or by subtracting half the penalty gradient from the estimating function (or equivalently, adding the gradient of the log partial prior to that function).

## 5. DISCUSSION

The present paper has addressed settings in which target models or parameters are not identified, and hence, the data cannot tell us whether we are close to or far from the target, even probabilistically. There are two sound responses by the analyst. One is to focus on describing the study and the data, resisting pressures to make inferences, in recognition that a single observational study will provide a basis for action only in extraordinary circumstances (Greenland, Gago-Dominguez and Castellao, 2004). If instead inference is mandated, as in pooled analyses to advise policy, we must admit we can only propose models that incorporate or are at least consistent with facts as we know them, and that all inferences are completely dependent on these modeling choices (including nonparametric or semi-parametric inferences).

In the latter process, we must recognize that there will always be an infinite number of such models and they will not all yield similar inferences. In this sense, statistical modeling provides only inferential possibilities rather than inferences. Any analysis should thus be viewed a part of a sensitivity analysis which depends on external plausibility considerations to reach conclusions (Greenland, 2005b; Vansteelandt et al., 2006). Results from single models are merely examples of what might be plausibly inferred, although just one plausible inference may suffice to demonstrate inherent limitations of the data.

Vansteelandt et al. (2006) offer a rationale for their region-constraint approach beyond those mentioned above (Section 2.1): To keep *ignorance* about $\boldsymbol{\theta}$ (uncertainty about bias), expressed as the region $\mathbf{R}$, distinct from *imprecision* (statistical or random error) as sources of uncertainty about $\boldsymbol{\tau}$. As shown in the example, the same distinction can be made when using relaxation penalties, and the two sources of uncertainty can be compared. Nonetheless, in observational health and social science there is no objective basis for the data model $G(\mathbf{a}; \boldsymbol{\gamma}, \boldsymbol{\theta})$ (no known randomizer, random sampler or physical law), which undermines the physical distinction between ignorance and imprecision. In these settings, $G(\mathbf{a}; \boldsymbol{\gamma}, \boldsymbol{\theta})$ merely expresses our conditional (residual) ignorance about where the data would fall even if we were given $(\boldsymbol{\gamma}, \boldsymbol{\theta})$; it differs from $H(\boldsymbol{\theta}; \boldsymbol{\lambda})$ only in that $G(\mathbf{a}; \boldsymbol{\gamma}, \boldsymbol{\theta})$ is invariably a conventional (intersubjective) form representing constraints that would have been enforced by an experimental design, but in reality were not enforced.

Whatever their value for summarization, conventional models do not satisfy plausibility considerations because they incorporate point constraints on unknown parameters. These include many bias parameters that can be forced to their null by successful design strategies, but are probably not null in most observational settings. Likewise, interval constraints rarely satisfy all plausibility considerations and thus may not be suitable for assessing total uncertainty (as opposed to providing sensitivity-analysis summaries). In contrast, relaxation penalties and priors allow expansion of conventional models and point constraints into the plausible realm, and thus can provide more plausible inferences. These capabilities justify their addition to basic statistical training for observational sciences. Progress beyond such penalties can be made only by obtaining data from a design that eliminates or at least partially identifies at least one previously nonidentified bias parameter (Rosenbaum, 1999; Greenland, 2005a).

## ACKNOWLEDGMENTS

I am grateful to Paul Gustafson and the referees for comments that greatly aided the revision.